# Atmospheric ionization by high-fluence, hard spectrum solar proton events and their probable appearance in the ice core archive

Adrian L. Melott[1], Brian C. Thomas[2], Claude M. Laird[3], Ben Neuenswander[4], and Dimitra Atri[5]

[1] Department of Physics and Astronomy, University of Kansas, Lawrence, Kansas 66045 USA
email melott@ku.edu

[2] Department of Physics and Astronomy, Washburn University, Topeka, Kansas 66621 USA
email brian.thomas@washburn.edu

[3] Retired.   email claude.m.laird@gmail.com

[4] Specialized Chemistry Center, University of Kansas, Lawrence, Kansas 66047 USA email
bennder@ku.edu

[5] Blue Marble Space Institute of Science, Seattle, Washington 98109 USA email
dimitra@bmsis.org

Running Title: Solar proton record in ice

Main Points:

1. Solar proton events produce nitrate visible in ice cores.
2. Air shower secondaries must be followed to produce correct results.
3. Nitrate deposition depends on spectral hardness.

**Abstract**

Solar energetic particles ionize the atmosphere, leading to production of nitrogen oxides. It has been suggested that some such events are visible as layers of nitrate in ice cores, yielding archives of energetic, high fluence solar proton events (SPEs). There has been controversy, due to slowness of transport for these species down from the upper stratosphere; past numerical simulations based on an analytic calculation have shown very little ionization below the mid stratosphere. These simulations suffer from deficiencies: they consider only soft SPEs and narrow energy ranges; spectral fits are poorly chosen; with few exceptions secondary particles in air showers are ignored. Using improved simulations that follow development of the proton-induced air shower, we find consistency with recent experiments showing substantial excess ionization down to 5 km. We compute nitrate available from the 23 February 1956 SPE, which had a high fluence, hard spectrum, and well-resolved associated nitrate peak in a Greenland ice core. For the first time, we find this event can account for ice core data with timely (~ 2 months) transport downward between 46 km and the surface, thus indicating an archive of high fluence, hard spectrum SPE covering the last several millennia. We discuss interpretations of this result, as well as the lack of a clearly-defined nitrate spike associated with the soft-spectrum 3-4 August 1972 SPE. We suggest that hard-spectrum SPEs, especially in the 6 months of polar winter, are detectable in ice cores, and that more work needs to be done to investigate this.

# 1. Introduction

## 1.1 General issues

Ionizing events in the Earth's atmosphere result in the formation of a variety of oxides of nitrogen [*Crutzen et al.*, 1975; *Heath et al.*, 1977; *Reid and McAfee,* 1978; *Reagan et al.*, 1981; *Jackman and McPeters*, 1985; *Jackman et al.*, 2000, 2001; *Thomas et al.,* 2005; *Thomas et al.*, 2007]. Many short-term variations in this ionization are dominated by Solar Proton Events (SPEs), which are of interest for understanding the physics of the Sun, possible hazards to our terrestrial electromagnetic technology, and even to spacecraft systems or crews, particularly on missions outside the Earth's magnetosphere. For this reason, it is valuable to have additional information on their rate of occurrence and intensity over longer time periods than available from direct cosmic ray monitoring measurements beginning in 1932 [*Shea and Smart*, 2000].

There has been a controversy over the possibility of using long-term ice deposits, such as those in Greenland or Antarctica, as an archive of solar activity. In particular, the concentration of nitrate as a function of depth in ice cores presents the possibility of identifying times of high ionization, as the main mechanism for removal of the nitrate from the atmosphere is precipitation in snow or rain. There have been competing claims over whether there is or can be a correspondence between ionization events and nitrate concentration in this archive [e.g. *Zeller and Parker,* 1981; *Legrand and Kirchner,* 1990*; Dreschhoff and Zeller*, 1998; *Wolff et al.,* 2012; *Smart et al.,* 2014; *Duderstadt et al.*, 2016; *Wolff et al.*, 2016; *Smart et al.,* 2016]. However, there is growing and now strong evidence for long-term modulation of polar nitrate deposition by solar activity [e.g. *Zeller and Parker,* 1981; *Palmer et al.*., 2001; *de Zafra et al.*, 2003; *Traversi et al*., 2012; *Poluianov et al.,* 2014; *Traversi et al*., 2016].

Nevertheless, the case for short term nitrate features (spikes) in the polar record from individual SPEs continues to be contentiously disputed. One of the important issues has concerned the argument that transport from the upper stratosphere and above (where the nitrate has been said to be formed) to the lower stratosphere and troposphere (where it can be precipitated out) is too slow for the events, which typically last a few days, to be detected [e.g. *Legrand and Kirchner,* 1990; *Duderstadt et al.*, 2014, hereafter D14].

## 1.2 Previous modeling

Some computations have shown nitrate formation only in the middle stratosphere and up. *Seppälä et al.* [2008] looked at hard-spectrum SPEs. They do not discuss their ionization modeling, but list *Verronen et al.* [2006] as the source of their methodology; this paper makes it clear that they use forcing by primary protons below 300 MeV only, and do not consider a full air shower with all the secondary particle types that are generated. As we shall see, this is no doubt why they found little ionization below 35 km. Another example is D14, who selected a relatively soft spectrum (i.e. no Ground Level Enhancement of neutrons or GLE) SPE on 9 November 2000 and considered only protons in a limited energy range of 10 - 300 MeV. D14 also used an exponential fit (based on data only up to 100 MeV) to the proton spectrum.  However, *Atwell et al.* [2011] have shown that an exponential fit may underestimate the correct ionization at the low energy end (10 MeV) by a factor of 2-3 and at the high energy end (300 MeV) by an order of magnitude.  Also, importantly, in the air shower modeling, apparently only the primary protons were included, which would not include ionization from secondaries that are responsible for effects in the lower atmosphere. All of the choices of D14 described here would discourage nitrate precipitation due to fewer ionizations and less penetration into the lower atmosphere.

After this paper had been initially submitted we were notified of *Duderstadt et al*. [2016], hereafter D16. Their method of doing air shower computations has been corrected relative to D14, and now appears to be in basic agreement with the more complete approach we have been advocating [*Laird et al.*, 2014] and using [*Atri and Melott*, 2011]. D16 use a chemical transport model (WACCM) to calculate nitrates, while we assume that every $NO_x$ ($NO_x = NO + NO_2$) molecule created through SPE ionization is added to the atmospheric column. Our nitrate values are an upper limit since we assume all $NO_x$ produced is converted to $HNO_3$ and then deposited. This is a valid assumption for altitudes below the upper stratosphere [*Calisto et al.* 2013]; other removal processes, primarily photolysis, are only important above about 40 km and not in polar winter, and also of little concern for rapid deposition in snow/ice.

However, there still are areas of concern:

(1) D16 do not cite *Nicoll and Harrison* [2014], an important experimental study whose title we state here for emphasis: "Detection of Lower Tropospheric Responses to Solar Energetic Particles at Midlatitudes". This study shows an ionization peak due to the 11 April 2013 SPE at 18 km and active increased ionization down to 5 km. We are able to closely match the ionization profile, as shown later. None of the events modeled by  D16 show these characteristics, indicating, instead, peak ionization near 50 km or above for most events they studied, and little to no ionization below 20 km, with the notable exceptions of their "hypothetical" hard and soft events based on the February 1956 and August 1972 SPEs.

(2) Despite basic agreement between D16's method and ours, two differences are important. As mentioned, one is the choice of events. The many simulations shown in D16 do not specifically include the real 1972 or 1956 high fluence events, instead they focus on hypothetical events or on those (such as 20 Jan 2005) which have no observed nitrate spike. The 1956 nitrate spike, which we address later, can be seen in *Smart et al.* [2014].

(3) The second and more fundamentally important difference is that we examine the total amount of nitrate expected to be produced in the air column and compare that with ice cores. The purpose of our study is to evaluate the additional amount of nitrate expected to be deposited following an SPE. This approach works very well for 1956, a well-known event.

On the other hand, comparing nitrate production from an SPE to the existing $NO_y$ atmospheric background at any given time does not achieve this goal. D16 define SPE produced $NO_y$ as $NO_y = N + NO + NO_2 + NO_3 + 2N_2O_5 + HNO_3 + HO_2NO_2 + ClONO_2 + BrONO_2$, but they then compare the amount of SPE-produced nitrate with the *total* $NO_y$ present in their WACCM model atmospheric reservoir, which includes many more species than they list in their definition. Calculating a percentage increase relative to this total $NO_y$ background to estimate potential nitrate deposition from SPEs is inappropriate for the following reasons:

(a) As documented [*Ridley et al.,* 2000 and references therein; *Jones et al.,* 2011], the atmospheric budget of nitrate can be dominated by organic nitrates, which have a long residence time. Nitrates in surface ice at the sites studied were found to be more closely correlated with nitric acid, which is expected to be produced by only a few sources, including SPEs, and which is deposited much more rapidly than organic nitrates. D16's total $NO_y$ background includes species such as PAN ($C_2H_3NO_5$) and other organic nitrates [e.g. *Kelly et al.*, 1967; *Jones et al.*, 1999; *Fahey et al.*, 1985; *Penkett et al.*, 2009]. In addition, D16 use the WACCM model, normalized for 2004. As documented in *Lamarque et al.* [2012] and tested in *Brakebusch et al.* [2013], this normalized model also includes organic nitrates from anthropgenic sources of $NO_y$. Since organic nitrates typically come out slowly, they are not an appropriate "background" against which to compare SPE produced nitrates. This background reservoir will be much larger than the $NO_y$ that is available to precipitate rapidly.

(b) It is well-known that polluted air masses, due primarily to anthropogenic $NO_x$ emissions, have significantly increased the Arctic $NO_y$ background, as evidenced by a doubling in surface nitrate deposition in Greenland since 1950, [*Herron,* 1982; *Mayewski et al.* 1986; *Fischer et al.*, 1998]. As the WACCM model adopted by D16 is set to 2004 [*Brakebusch et al,* 2013; *Lamarque et al.,* 2012], it uses a much larger $NO_y$ background or reservoir than actually existed during the early and prehistoric SPEs that are of interest.

D16's choice to include organic and other anthropogenic sources of $NO_y$ has consequently both severely inflated the $NO_y$ background beyond what should be considered for evaluating past SPE impacts and significantly reduced the apparent relative production of $NO_y$ by SPEs. Their analysis makes it appear that any resulting nitrate deposition, even from large, scaled-up events, would be too small to detect above background sources when a more straightforward approach suggests differently. In contrast, our method of directly computing the absolute amount of additional nitrate available for deposition is more closely tied to the ground-truth nitrate measurements in ice cores.

## 1.3 Directly measured data on atmospheric ionization

As mentioned above, recent, balloon-borne, direct measurements [*Nicoll and Harrison*, 2014] show substantial ionization from an SPE within the lower stratosphere and troposphere. Considerable full modeling of air showers with proton primaries has been completed, which takes full account of ionization caused by the secondaries in the air shower [*Atri et* al., 2010; *Usoskin et al.,* 2006*; Usoskin et al.,* 2010; *Usoskin et al.,* 2011]. This is the basis of our work here. In what follows, we briefly compare these ionization results with older computations [*Jackman et al.,* 1980; D14]. Then, we use our results to estimate the nitrate production as a function of altitude for the major SPEs of 23 February 1956 and 3-4 August 1972, and compare the results with experimental data from ice cores.

## 2. Numerical modeling

## 2.1 Numerical procedure determining ionization from primary cosmic rays

In this work we use existing results of computations of atmospheric ionization. We employ two data sets, one a result of our own computations, the other provided by I. Usoskin [personal communication, 2015]. The calculations provided by I. Usoskin were performed with a hybrid model that uses an analytic approach for proton energies between 10 MeV and 100 MeV, combined with results of full air shower modeling for energies above 100 MeV, to a maximum of 10 GeV. The low energy calculations follow the approach described in *Jackman et al.* [1980] (see also *Verronen et al.* [2005], *Jackman et al.* [2011], and *Kokorowski et al.* [2012]). The energy deposited in altitude bin $i$ by protons with kinetic energy $E$ and pitch angle $\theta$ is given by

$$E_{di}(\theta, E) = E - \left\{ -\frac{\Delta Z_i}{A} \sec\theta + E^B \right\}^{1/B} \text{MeV}$$

using the range energy relation [*Bethe and Ashkin*, 1953; *Whaling*, 1958; *Sternheimer*, 1959; *Green and Peterson*, 1968]:

$$R(E) = A \left( \frac{E}{1 \text{ MeV}} \right)^B \text{gm cm}^{-2}$$

with $A = 2.71 \times 10^{-3}$ and $B = 1.72$ for $1 \leq E \leq 70$ MeV.

For proton energies above 100 MeV, ionization values use results of numerical modeling performed with the CORSIKA package, a widely used Monte Carlo based simulation tool [*Heck et al.*, 1998]. This combined approach makes the model more accurate than simple models that use only the Bethe-Bloch equation for Bragg curve calculations. The model is implemented with protons as primary particles incident isotropically from the hemisphere and setting the secondary particle cutoff energies as described below.

CORSIKA is calibrated with state of the art experimental results worldwide. The model is capable of simulating all the electromagnetic and hadronic interactions resulting from a particle-induced cascade in the atmosphere down to the cutoff energy possible with the model. The lower energy cutoffs are as follows: hadrons = 50 MeV, muons = 50 MeV, electrons = 50 keV, and photons = 50 keV. The CORSIKA code allows the use of two hadronic interaction models, one for interactions up to 80 GeV, and the other one beyond that. Since all the interactions modeled here are below 80 GeV, the outcome depends only on the low energy model in CORSIKA. The data provided by I. Usoskin were produced using the FLUKA (v.2006.3b) [*Fassò et al.*, 2001] model in CORSIKA. Full details on that implementation may be found in *Usoskin et al.* [2006], *Usoskin et al.* [2010], and *Usoskin et al.* [2011].

Our own calculations were performed for proton energies above 300 MeV, and we used the UrQMD (Ultrarelativistic Quantum Molecular Dynamics) model in CORSIKA [*Bass et al.,* 1998; *Bleicher et al.,* 1999], which is used widely for air shower simulations. We also used this model in our earlier work to compute atmospheric ionization with higher energy primaries [*Atri et al.,* 2010]. Showers are tracked from the point of first interaction in the upper atmosphere down to the ground level in bins of 10 g cm$^{-2}$. The standard bin size recommended for simulations is 20 g cm$^{-2}$ [*Heck and Pierog*, 2000]; we chose 10 g cm$^{-2}$ for higher accuracy, but had to use $10^7$ showers for each energy to achieve satisfactory numerical and statistical accuracy. A link to the lookup table is provided in the paper where these results can be accessed.

## 2.2 Numerical procedure computing ionization and dissociation from specific events

The procedure used to compute ionization for solar protons is described in *Usoskin et al.* [2011], using the ionization tables described in the previous section. The total ionization for the event at altitude *z* is given by:

$$I(z) = \int_{T_1}^{T_2} S(T) \cdot Y(z, T) dT,$$

where $S(T)$ gives the number of protons with kinetic energy $T$, $Y(z,T)$ gives the number of ion pairs produced at altitude $h$ by a single proton with kinetic energy $T$, and $T_1$ and $T_2$ give the range of proton kinetic energy values.

For protons of rigidity $R$ in GV and kinetic energy $T$ in GeV, the event integrated omnidirectional integrated fluence (protons cm$^{-2}$; integrated over the event-specific duration) is given by the Band function:

$$J(>R) = J_0 \cdot R^{-\gamma_1} e^{-R/R_0}, \text{for } R \leq (\gamma_2 - \gamma_1)R_0,$$

$$J(>R) = J_0 \cdot A \cdot R^{-\gamma_2}, \text{for } R > (\gamma_2 - \gamma_1)R_0,$$

where

$$A = [(\gamma_2 - \gamma_1)R_0]^{(\gamma_2 - \gamma_1)} e^{(\gamma_1 - \gamma_2)},$$

$$R = \sqrt{T^2 - 2T_0 \cdot T},$$

and $T_0 = 0.938$ GeV is the rest mass energy of the proton. The numerical method uses the event integrated differential spectrum in proton kinetic energy (protons cm$^{-2}$ sr$^{-1}$ GeV$^{-1}$):

$$S = \frac{1}{4\pi} J_0 \cdot R^{-\gamma_1} e^{-R/R_0} \frac{(\gamma_1 R_0 + R)(T + T_0)}{R^2 R_0}, \text{for } R \leq (\gamma_2 - \gamma_1)R_0,$$

$$S = \frac{1}{4\pi} J_0 \cdot A \cdot \gamma_2 \cdot R^{-\gamma_2} \frac{T + T_0}{R^2}, \text{for } R > (\gamma_2 - \gamma_1)R_0.$$

For the 23 February 1956 event, the spectrum values are $J_0 = 1.747 \times 10^8, R_0 = 0.5661, \gamma_1 = 1.758$, and $\gamma_2 = 5.04$ . For the 3-4 August 1972 SPE the values are $J_0 = 6.340 \times 10^6, R_0 = 0.2980, \gamma_1 = 3.260$, and $\gamma_2 = 6.27$; both events [*Tylka and Dietrich*, 2009].

## 2.3 Comparison with other work, data sources, and test cases

*Jackman et al.* [1980] follow only ionization produced by primaries and secondary electrons generated by them. *Jackman et al.* [2011] apparently supplemented this with data inferred from GLEs, usually a burst of excess neutrons observed at the ground [*Overholt et al.,* 2013], but did not include muons, which dominate ionization at lower altitudes [*Atri and Melott,* 2011]. A more complete approach [*Bazilevskaya et al.,* 2008; *Atri et al.,* 2010; *Usoskin et al.,* 2011; *Calisto et* *al.,* 2013] follows the production of all secondaries. *Atri et al.* [2010] published lookup tables so that new simulations do not need to be done for each new spectrum. Ionization tables as a function of altitude for primaries ranging from 70 MeV to 1 PeV are given. An analytical fit (described above) is available for primaries below 70 MeV [*Jackman et al.,* 1980].  Any SPE spectrum can be modeled by convolution with this table.

The general characteristics of the table of *Atri and Melott* [2010] can be compared with the table of *Usoskin and Kovaltsov* [2006] and *Usoskin et al.* [2010]. The latter includes results for alpha particles, which *Atri and Melott* does not. *Atri and Melott* shows ten entries per decade in primary energy, while *Usoskin and Kovaltsov* [2006] and *Usoskin et al.* [2010] show two. *Atri and Melott* shows 46 vertical bins in atmospheric density; *Usoskin and Kovaltso*v [2006] and *Usoskin et al.* [2010] combined show 52.  *Usoskin and Kovaltso*v [2006] and *Usoskin et al.* [2010] combined show primary energies from 100 MeV up to one TeV; *Atri and Melott*  includes primary energies from 300 MeV up to one PeV (three orders of magnitude higher than the other tables). One PeV is far beyond the range necessary for SPE events, but does assure that the high energy tail will be included, which is important for hard, major GLE producing events as well as making possible the modeling of effects of nearby supernovae, gamma-ray bursts or other high-energy astrophysical events [*Melott and Thomas,* 2011*; Piran and Jimenez,* 2014]. As noted earlier, the two tables agree well in their region of overlap.

As a check on our modeling results, we have compared them to an ionization profile for the 20 January 2005 SPE reported in *Usoskin et al.* [2011].  Figure 1 shows results of our model applied to the 20 January 2005 SPE case (assuming a 1 day duration) along with those from Figure 2B of *Usoskin et al.* [2011] (for the same SPE).  We note that the agreement is very good below ~23km.  A divergence does begin to appear at the highest altitudes, probably because *Usoskin et al.* [2011] included lower energy primaries, which contribute more ionization at higher altitudes.

Both sets of results include ionization far below 20 km, the cutoff shown in D14. The GLE on 20 Jan 2005 clearly shows that ionization in the troposphere is possible, even though not shown in the simulations of D14, who considered a softer event whose conclusions cannot be generalized to harder spectrum events.  However, the production of a significant flux of muons for detection on the ground and ionization in the lower atmosphere requires some primaries about 10 GeV or above [*Atri and Melott* 2011], while a GLE (detection of a neutron excess on the ground) can be generated by primary particles with energy below 2 GeV [see, e.g. *Overholt et al.* 2014]. A GLE is therefore a necessary, but not sufficient condition for substantial lower-atmosphere effects.

In Figure 2 we compare ionization rates computed using a full air-shower treatment (provided by I. Usoskin) versus ionization rates computed using only the low-energy analytic method described in section 2.1 (data provided by C. Jackman, downloaded from http://solarisheppa.geomar.de/solarprotonfluxes). We show ionization rates for two SPEs; 3-4 August 1972, which had a soft spectrum (but high fluence), and 20 January 2005, which had a hard spectrum (but lower fluence). As can easily be seen, the analytic-only method significantly underestimates the ionization in the mid to low atmosphere, even for the soft spectrum case (1972). In fact, the low energy analytic-only method yields zero ionization below the stratosphere in both cases, while the air-shower method shows significant ionization in the troposphere.

We supplemented the above test with comparison against the balloon data of *Nicoll and Harrison* [2014], hereafter NH, which directly measured ionization from the 11 April 2013 SPE over a range of altitudes from 1 to 30 km. This comparison is shown in Figure 3. We note that in this experiment the SPE-induced ionization, as shown by NH, peaked at about 18 km, and is clearly nonzero down to 5 km. The GOES high energy proton and alpha data provide the flux and spectral information on this event for energies in excess of 433 MeV/nucleon, and we found that by implementing a low-flux, power-law index spectrum of $dN/dE = 0.91xE^{-2.4}$ protons (cm$^{-2}$ sr$^{-1}$ GeV$^{-1}$ s$^{-1}$) consistent with these data we reproduce the ionization plotted in NH. (Due to the geomagnetic cutoff rigidity 47.5° N in the vicinity of the UK [*Smart et al.,* 2006], the lower end of the spectrum is irrelevant here.) The GOES proton data are on the NOAA database http://satdat.ngdc.noaa.gov/sem/goes/data/

As we discuss later, winter conditions are generally much better than summer for the preservation of ice core nitrate signals. Since fine resolution winter ice core data from Greenland are not available for 2005 or 2013 we cannot do a direct comparison with nitrate production for either event. We therefore now focus on the SPE of 23 February 1956, which is the largest GLE in amplitude (a 46-fold increase in 15-minute data) and fluence above 200 MeV in the cosmic ray monitoring history [*Kovaltsov et al.,* 2014], and had a hard spectrum [*Pfotzer,* 1958; *Smart and Shea,* 1990; *Belov et al.,* 2005]. This event was world-wide in character and observed at the geomagnetic equator by [*Sarabhai et al.,* 1956], who estimated the maximum energy of solar protons to be >50 GeV. More importantly, since our goal is to compare atmospheric computations with ice core data, there is an observed impulsive nitrate event in the ice core from Summit, Greenland [*Smart et al.,* 2014, 2016]. In this case quantitative comparison is possible proceeding from the measured spectrum to the amount of nitrate produced in the atmosphere and the amount found in the ice (see Figure 4A). We use data provided by I. Usoskin [personal communication, 2015] which was produced following the procedure described in sections 2.1 and 2.2, using the derived high energy fluence spectrum specified for 1956 in the previous section of this paper [*Tylka and Dietrich,* 2009] over the range 10 MeV to 10 GeV, although there were higher energy primaries present, so our conclusions are conservative in that they will underestimate the amount of ionization and its penetration into the troposphere.

In addition, we examine the large SPE of 3-4 August 1972, which had a high fluence but a very soft spectrum. The purpose of this is primarily to demonstrate the ionization changes due to a soft spectrum.

## 3. Comparison with ice core data for 1956

In the following, we compare the results of our computations with the GISP2H ice core data [*Dreschhoff and Zeller*, 1998], which show a nitrate spike coincident with the SPE and its accompanying GLE (Figure 4A). The annual cycle of summer high - winter low can be observed. The nitrate deposition above background for the winter 1956 GISP2H peak identified by *McCracken et al.* [2001] that is superimposed on the annual cycle was determined by computing the average concentration of the 4 data points that comprise the peak and subtracting from it the local background represented by the average of the two adjacent local minima. The resulting average nitrate concentration (33.9 ng $g^{-1}$) above background was then multiplied by the thickness of all four samples (6.0 cm) and by the average density of the ice at that depth (0.568 g $cm^{-3}$), yielding a deposition of 116 ng $NO_3$ $cm^{-2}$. In Sections 4 and 5 we will specifically investigate whether the SPE of 23 February 1956 can provide this amount of $NO_3$.

It is difficult to put an error estimate on this number because there are many potential sources of error related to deposition that are external or environmental. The calculated nitrate deposition was derived from the best available data. The nitrate concentration instrument error is less than a few percent and the density measurement error is roughly a few percent. Additional cores and measurements will be required to better estimate errors resulting from other environmental factors.

The Figure 4 horizontal axes were constructed under the assumption of linearity between time and deposition within a given year (i.e. equally-spaced samples). This time approximation is not strictly true at the sub annual scale . *Dibb and Fahnestock* [2004] conducted a 2-year snow stake study of accumulation rates at Summit. They concluded that most of the accumulation occurs in summer (warm) months. Using their monthly averages for those 2 years to refine our approximation starting on March 1, we estimate that it took roughly 2 - 4 months for the winter 1956 nitrate peak to be deposited. Increased photochemistry during the summer and absence of the polar vortex may make it more difficult to see an SPE-produced nitrate peak then. So, events like this SPE may need to be large and have a bias toward winter occurrence in order to be observed in nitrates.

## 4. Results

In Figure 5 we show a summary of the atmospheric ionization results for 23 February 1956 and 3-4 August 1972, obtained from I. Usoskin [personal communication, 2015], generated using the procedure detailed in Section 2. The dashed lines show the background ionization due to Galactic Cosmic Rays (GCRs) in February 1956 and August 1972, and the solid lines the ionization rate produced by the SPEs. The proton spectrum used is event-integrated but here we present ionization rate, assuming a 1 day duration for the SPEs, in order to allow for comparison with GCR ionization rates acquired from http://cosmicrays.oulu.fi/CRII/CRII.html [*Usoskin and Kovaltsov,* 2006; *Usoskin et al.,* 2010].

The ionization values are mapped into nitrate production as follows. Ionization and dissociation of $N_2$ and $O_2$ by solar protons leads to production of about 1.25 N atoms per ion pair [*Porter et al.,* 1976; *Jackman et al.,* 2005]. These N atoms are distributed between the electronic ground state $N(^4S)$ and the excited state $N(^2D)$. The excited state $N(^2D)$ production determines the net production of $NO_x$. It is usually assumed that 55% of the N atoms are in the excited state [*Rusch et al.,* 1981; *Jackman et al.,* 2005]; however, the actual value may range from 15% to 95% [*Funke et al.,* 2011; *Sinnhuber et al.,* 2012]. Here we use 55%, which in any case is about the middle of that range. In order to get an estimate for the total $NO_3$ deposited we first convert the total ionization for the entire event at each altitude bin in the model to $NO_x$ produced by multiplying by 1.25 x 0.55. In the stratosphere, the $NO_x$ produced is subsequently converted to $HNO_3$ over a few weeks' time [*Funke et al.,* 2011]. Therefore, we take the $NO_x$ value computed at each altitude, assume it is all converted to $HNO_3$, and then sum over a given altitude range to give a total event-integrated deposited value in ng $NO_3$ cm$^{-2}$. It is important to emphasize that our ionization results used to compute nitrate deposition represent the total ionization for the event considered; that is, integrated over the entire event duration. This time integration is implicit in the proton spectrum parameters used in this work, so that when we apply the spectrum in our ionization model we have computed ionization for the event as a whole, rather than an ionization rate, which then leads to a total nitrate deposition value.

In Figure 6 we present results of these calculations for the 23 February 1956 and 3-4 August 1972 events, along with the corresponding results due to GCR ionization at these two dates. GCR ionization produces nitrate in the same way SPEs do, and so represents a background source with the same properties (as opposed to other background sources such as $NO_y$ that is transported from lower latitudes). Nitrate values (in ng $NO_3$ cm$^{-2}$) summed from the ground to the altitude indicated are given, computed from modeled ionization as described above. We also show the difference between the SPE and GCR values for both cases. We are interested here in the enhancement due to SPEs above the background GCR ionization. Therefore, assuming a normal GCR background, the enhancement of nitrate for deposition is given by the SPE-GCR values. On the other hand, GLEs are often associated with a reduction in GCR flux known as a Forbush decrease. In the case of a complete elimination of GCR background (an extreme example), the enhancement is given by the SPE values alone. This allows us to define a maximum (SPE values only) and minimum (SPE-GCR) enhancement.

We find that summing from the surface (3.2 km at Summit) to an altitude of 44 - 46 km, we reach approximately the 116 ng $NO_3$ cm$^{-2}$ found in the ice core peak (see Figures 4A and 6), both with and without the GCR background (and hence the presence/absence of a Forbush decrease does not make a significant difference for our results). We also note that half the needed production occurs below 30 km. So, there is sufficient nitrate produced in the atmospheric column to account for this ice core peak contemporaneous with the long duration, high energy, hard spectrum SPE, provided timely transport downward from as high as 46 km is feasible.

As noted above, there is uncertainty in the fraction of N atoms in the excited state following ionization, which determines the net production of $NO_3$ that is eventually precipitated. If we take the extreme maximum fraction (95%), summing to 32 km gives approximately the ice core value. On the other hand, if we take the extreme minimum fraction (15%), there is not enough $NO_3$ produced over the entire atmospheric column to explain the ice core peak. As discussed above, we take 55% as the best estimate, but note here that the deposited nitrate will depend strongly on the exact conversion fraction used.

We contrast this with the results from the 3-4 August 1972 SPE, which are shown in Figure 5. It produced only slightly less nitrate when summed up to 100 km (Figure 6), but the distribution was very different, emphasizing ionization at higher altitudes. In contrast to 1956, one would have to go to 68 km to account for an ice core spike like 1956 from the 1972 results.

Our results constitute the first comparison of ionization and nitrate production by high fluence SPEs with nitrate deposits in ice cores, which are candidate signals of these events. We find that at least for a major hard spectrum event, there is good agreement between the computed results and the amount measured in the ice core.

## 5. Discussion

### 5.1 Was sufficient nitrate available from the 23 February 1956 SPE for prompt transport?

As can be seen (Figure 6), although sufficient nitrate is produced in the stratosphere and below, tropospheric nitrate (which could be deposited very rapidly) by itself is insufficient to account for the nitrate peak in the ice core data for 1956. The data appear to require inclusion from the mid-stratosphere downward, still very different from past assertions that nitrates are made from the mid-stratosphere *upward* [e.g. *Wolff et al.*, 2012; D14]. We see three possibilities: (1) Association of the peak with the SPE is simply incorrect. (2) The SPE has a much harder spectrum than assumed, somehow turning back up at higher energies. A harder spectrum would deposit more energy in the troposphere. (3) Transport downward in the mid-stratosphere can happen rapidly enough to account for this peak—and even narrower peaks identified with GLEs in the 1940s [*Smart et al.* 2014, 2016]. We discuss each of these possibilities for resolving the question.

### 5.2 Possible misinterpretation of coincidence between 23 Feb 1956 event and nitrate peak

A number of studies beginning with *Wilson and House* [1965] have attempted to demonstrate and refine the association between solar activity and nitrate variability in polar ice cores [e.g. *Zeller and Parker*, 1981; *Laird et al.,* 1982; *Dreschhoff and Zeller,* 1994, 1998; *McCracken et al.,* 2001 and *Smart et al*., 2014, 2016]. These studies have invoked statistical correlations between some of the impulsive nitrate peaks that appear to be contemporaneous with known large SPEs. However, spurious correlations could occur in about four ways:

(1) The apparent coincidence in timing could be spurious. Ice core dating in historical times is based on identifying major, absolutely-dated volcanic eruptions in the sulfate and conductivity (determined primarily by sulfate for the anions) records, and interpolating between these tie points using impurity records that display annual cycles. The greater the distance between volcanic tie points, the less certain the dating at the annual scale. The GISP2H conductivity record used the Hekla (1947) and Askja (1961) Iceland eruptions as absolutely dated tie points, together with counting and interpolation of annual nitrate cycles in between, to determine 1956 as the most probable year for the spike in question. However, this peak possibly could be misdated still by as much as one or, less likely, two years.

(2) Post depositional processing at the surface including physical mixing by winds, sublimation, condensation and chemical processes involving firn-air interface exchange [e.g. *Laird et al.,* 1987*; Dibb and Jaffrezo,* 1997; D14] tend to reduce the highest nitrate concentrations found at the surface as they are incorporated into the ice column. Still, it appears feasible for high summer surface nitrate levels unrelated to SPEs to be incorporated *in toto* on occasion as isolated spikes at random points across the ice sheet and at least partly preserved at depth [e.g. *Laird et al.,* 1987; *Dibb et al.,* 2007]. Without additional fine resolution nitrate studies of ice cores there is no way to verify whether this effect is responsible for the 1956 spike in the GISP2H core, or whether the spike persists across the ice sheet (which seems more likely, given its apparent occurrence during winter).

(3) Downward transport from the stratosphere (due to subsidence in the winter polar vortex) in lieu of an SPE is a likely background source of nitrate in ice cores. However, as with the anthropogenic background source in the troposphere (discussed below), this process is generally considered to be too slow and steady [e.g. *Legrand and Kirchner,* 1990; D14] to produce the sharp nitrate spikes observed in the GISP2H core. The 1956 spike was only a 4% enhancement above the integrated background for the entire year, so slow, steady deposition of this excess nitrate would be very difficult to identify.

(4) Nitrate peaks have been associated with other, intermittent sources including (i) air plumes with anthropogenic pollution [*Herron,* 1982; *Mayewski et al.,* 1990], (ii) biomass burning [*Whitlow et al.,* 1994] and (iii) sea salt and possibly dust particulate intrusions [*Dibb and Jaffrezo*, 1997; *Wolff et al.,* 2008]. These potential sources of interference with nitrate are the ones that have been invoked most recently to argue against a possible SPE source (e.g. *Wolff et al.* [2012] and D14).

Anthropogenic sources, due to the combustion of fossil fuels and production and application of fertilizer, are observed in the nitrate record increasingly since 1950 as a general upward trend over the last 60 - 70 years and may contribute to the annual cycle background, but are unlikely to produce an impulsive winter nitrate spike of 2-4 month duration. Biomass burning, primarily from large forest fires, which can produce ammonium nitrate but is primarily a summer phenomenon, is also unlikely to be the source of the 1956 GISP2H winter nitrate peak. Summit cores such as GISP2 [*P. A. Mayewski,* personal communication, 2015] and Zoe [*J. R. McConnell*, personal communication, 2012] have been analyzed for a suite of chemical species, but either they show no interference from biomass burning in the winter of 1956 or lack the resolution required for comparison with GISP2H [*Smart et al.,* 2014, 2016]. Similarly, nitrate salt from marine influences is unlikely to exert much influence at the remote, high-elevation, Summit Station [*Wolff et al.,* 2008].

Of course the winter argument alone does not disprove interference by these nitrate sources in the form of aerosols such as from biomass burning, fertilizer and sea salts. There is, however, additional information in the GISP2H data set that can be used to get a handle on the potential source of the nitrate spike in question, namely the conductivity measurements. To do this analysis, we assumed a constant background from the only other major anion, sulfate. This seems reasonable given that no significant volcanic activity is reported during this time period. Standard molar conductivity tables for dilute solutions [*Haynes*, 2015] predict a rise in the conductivity level associated with the 4-point, averaged, 1956 nitrate spike (Figure 4A) of about 0.230 μS cm$^{-1}$ if the nitrate is acidic, as expected if precipitated from PSCs, and 0.066 - 0.079 μS cm$^{-1}$ if the peak is due to an aerosol of nitrate salt (i.e. ammonium, sodium, magnesium, or calcium nitrate), as expected if from biomass burning or marine sources. The actual rise above background of the integrated, colocated, 4-point, 1956 GISP2 H conductivity is 0.205 μS cm$^{-1}$. This is only 11% less than predicted for a nitric acid source, but 159% - 212% more than predicted if the spike was formed by a nitrate salt. In principle, the 1956 peak could result from a combination of nitrate salts instead of nitric acid, but this is increasingly improbable as it requires multiple winter sources and ignores the most straight forward explanation. In lieu of further data, this result lends strong additional support for the conclusion that the 1956 winter nitrate spike is mostly acidic and therefore what would be expected from the Feb 1956 SPE.

### 5.3 Was the event spectrum harder?

Muons are important to understanding ionization in the troposphere. We must allow that SPEs might have much more complicated spectra with a very strong hard component that has not been well-measured. On the other hand, such an energetic component is contradicted by other studies [*Swinson and Shea,* 1990], which still indicate primaries up to 25 GeV and higher [*Sarabhai et al.,* 1956]. There are recent studies that could be explained by a possible anomalous hard component reaching the ground [*Overholt and Melott*, 2015]. However, we stress that this is a speculative idea.

### 5.4 Transport from 45 km is possible and sufficient for the data

In order to account for the 1956 impulsive ice core peak, the total odd nitrogen (i.e. nitrate and its precursors) found in the atmospheric column from the surface up to 44 - 46 km would have to be deposited over a period of roughly 2 - 4 months. The short timing of this deposition is an issue and has not been supported by previous modeling. However, since these altitudes are much lower than were previously thought, the mechanism of vertical subsidence due to the polar vortex [*Traub et al.*, 1995; *Vogel et al.*, 2008; D14] coupled with denitrification by PSCs [*Crutzen and Arnold*, 1986; *Fahey et al.*, 1990; *Hamill and Toon*, 1991; *Northway et al.*, 2002; *Solomon*, 1999; *Toon et al.*, 1986] and followed by rapid deposition to the surface becomes viable. Since half the nitrate required for the 1956 winter peak is produced below 30 km, it will be subject to the denitrification process and removal in PSCs within about one month, as described below

Within the winter polar vortex, subsidence rates and their estimates vary. For example, an average descent rate of roughly 0.19 km day$^{-1}$ for 32 - 49 km altitude can be inferred from Fig 1a in the *Vogel et al.* [2008] CLaMS model. A previous study by *Traub et al.* [1995] measured an average of 0.43 km day$^{-1}$ at 18 km altitude during the winter months of 1992. The D14 WACCM model propagated NO$_x$ at a rate of ~0.33 km day$^{-1}$ from the upper to middle stratosphere during the months of November - December. Using the intermediate rate from D14, a parcel of air between 30 - 46 km, containing the upper half of the NO$_y$ column that is required to produce the spike, would travel down to an altitude of PSC formation (28 km) within 8 weeks, and from 34 km (assuming 95% conversion efficiency from N($^2$D) to NO$_x$) in less than two weeks. Normal downward transport with a speed of 200-300 m/day would also transport NO$_y$ from 34-46 km down to 14 km within 8 - 23 weeks (2 - 5 months).

Although more variable and generally less intense than in the Antarctic, denitrification of the Arctic mid to low stratosphere can be significant [*Hamill and Toon*, 1991; *Mann et al.*, 2003; *Popp et al.*, 2001]. Denitrification occurs due to gravitational settling of PSCs (type I), which form at sufficiently low temperatures and consist of nitric acid trihydrate (NAT) particles, generally between an altitude of $15 - 25$ km [*Voight et al.*, 2000], although altitudes from 10-28 km are possible [*Hamill and Toon*, 1991]. While the altitude of PSC formation is generally lower than 28 km, the presence of a PSC is evidence that denitrification has already occurred at significantly higher altitudes since it takes time for NAT particles to grow, during which they have fallen vertically as far as ~6 km over a period of 6 days [*Fahey et al.*, 2001]. The altitude for denitrification, therefore, encompasses the 15 - 31 km range. The speed of downward vertical transport for NAT's is a function of the particle size; large particles ($10 - 20$ μm) are subject to downward gravitational transport at a rate of $1 - 2$ km/day which is sufficient to transport to the troposphere within $1 - 2$ weeks, provided temperatures remain low enough that the particles do not evaporate [*Fahey et al.*, 2001; *Hamill and Toon*, 1991; *Northway et al.*, 2002]. Smaller particles move downward more slowly. Aircraft have measured NAT particles of this size in the Arctic [*Fahey et al.*, 2001; *Northway et al.*, 2002b].

Once nitric acid crosses the tropopause, it is subject to the relatively short residence time of particles in the troposphere (3-10 days) [*Papastefanou*, 2006; *Poet et al.*, 1972; *Rodhe and Grandell*, 1972; *Warneck*, 1999; *Winkler et al.*, 1998] where it will be largely removed in a couple of snowstorms [*Schwikowski et al.,* 1998]. Although widespread PSC formation has been seen in the Arctic during winters with a strong polar vortex, it is more common for PSC's to be localized [*Engel et al.*, 2013; *Fahey et al.*, 2001; *Pawson et al.*, 1995]. Unfortunately, temperature data for the stratosphere above Summit is not available for 1956 and troposphere temperatures may not correlate very well with those of the stratosphere [*Liu and Schuurmans*, 1990]. We note, however, that the Arctic vortex has been weakening due to climate change and would have been stronger in the past.

The involvement of PSCs in nitrate deposition of the winter 1956 ice core peak is also supported by the conductivity analysis in section 5.2, which indicates an acid source. Therefore, subsidence and denitrification by PSCs can qualitatively account for the ~2-4 month time scale consistent with the 1956 GISP2H winter ice core peak, however, more realistic modeling is necessary to establish this quantitatively. Proper modeling of PSC development is lacking (e.g. D16) and is complicated by an incomplete understanding of NAT nucleation and evolution, including evidence suggesting that strong SPEs can enhance the formation of large NAT particles [*Engel et al.*, 2013; *Mironova et al.*, 2008; *Popp et al.*, 2006; *Yu*, 2004].

## 5.5 Computation results for the 3-4 August 1972 SPE

We found total nitrate production from the ground to 100 km for the 3-4 August 1972 SPE to be similar to that for 1956 (shown in Figures 5 and 6). However there were considerably smaller contributions at lower altitudes due to the softer spectrum.

There are conflicting claims for the observation of this event in the ice core records [e.g. *Laird et al.* 1982; *Zeller et al.*, 1986; *Legrand and Kirchner*, 1990; *Wolff et al.*, 2012], which could be more problematic, given that it was not a winter event in the northern hemisphere and would be subject to photolysis in the atmosphere and at the surface and other deposition issues, as well as an absent polar vortex. Observations in Antarctica could also be hampered by the fact that high deposition areas, such as occur at Summit, are generally absent on the high south polar plateau or are along the coast, where they are subject to marine interferences. We show this time period from the GISP2H ice core in Figure 4B. The event does not show up clearly as either a nitrate or conductivity spike above background fluctuations, but, interestingly, there is a second subdued nitrate peak, which occurs after the prominent summer maximum that might partly represent deposition from this softer SPE over a 4 - 6 month interval. If so, it would require $NO_y$ inputs from above 65 km, which seems to be problematic. Also, the corresponding conductivity local background is higher than the values associated with the nitrate peak making any assessment of source more difficult. We regard these ambiguous findings as due to a combination of the soft spectrum and the less than optimal conditions described above.

## 6. Conclusions

1. The results of full air shower simulation, including effects of secondary particles, are necessary to properly model atmospheric ionization by SPEs, otherwise lower atmosphere ionization is underestimated. It is now possible to bypass redoing the full air shower simulation step for each new event by using the tables published by *Atri et al.* [2010] or alternately by *Usoskin and Kovaltsov* [2006] and *Usoskin et al.* [2010]. See the comparison of the two approaches in section 2.3. Our estimates are conservative, in that they do not include alpha particles, which will increase ionization by about 10% [*Jackman,* 2013]. Primary electrons are unimportant except at very high altitudes.

2. Computations based on air showers with low fluence or a soft spectrum [e.g. D14] are not generalizable to high-fluence showers or those with hard primary energy spectra. In order to be done correctly at mid-stratospheric down to tropospheric levels, atmospheric ionization must include the effects of secondaries and must include the high-energy primaries, which penetrate to lower altitudes and produce effects there. It may be that only showers with fairly hard energy spectra and high fluence are traceable as elevated nitrate levels in the ice cores. The results of *Nicoll and Harrison* [2014] are perfectly reasonable on the basis of theory for a hard-spectrum, moderate fluence event, showing excess ionization down to 5 km. As research into the potential impact of SPEs on nitrate deposition in polar snow and ice has evolved, it has become increasingly evident that major events are required to produce the needed ionization.

As discussed in Section 1, D16 have improved on the methods used in D14 and indeed produce more ionization at lower altitude (attributable to having adopted the more correct air-shower ionization calculations that we have been advocating and use in this work), though they still do not address or cite the observed tropospheric ionization reported in *Nicoll and Harrison* [2014], which we reproduce here. While not irrelevant to our work, the D16 results are given in terms of percentage increase in $NO_y$ against the normal atmospheric background (see also *Calisto et al.,* 2013]). This answers a different question than we seek to answer here – that is, can SPEs produce enough nitrate (in absolute terms) available for rapid deposition to explain amounts observed in ice cores? As we have shown, the answer is yes in at least one case.

3. There may be a bias away from visibility of summer events due to increased photochemistry and absence of the polar vortex, but more needs to be done here. A higher fluence may be required in order to be visible in summer ice deposits. *Smart et al.* [2014] noted one summer event in an ice core associated with a hard spectrum GLE in the 1940's. This deficiency could be surmounted by data from both polar regions. Polar winters are about 6 months, which helps.

4. Events with hard spectrum GLEs are more likely to appear as a layer of elevated nitric acid in the ice that is associated with higher conductivities and precipitation from PSCs. A strong GLE indicates deep penetration into the atmosphere, suggesting a hard primary spectrum and the possibility of rapid deposition as indicated in *Smart et al.* [2014, 2016].

5. There are no direct data for the SPE spectrum or presence of a GLE associated with the Carrington Event, which occurred at the beginning of September 1859 [*Clauer and Siscoe*, 2006]. It is therefore unclear whether or not its effects are expected to appear in ice core data. *Wolff et al.* [2012] found no evidence for nitrates in ice cores, but *Smart et al.* [2014, 2016] showed that the resolution of the data used by *Wolff et al.* was too poor to detect nitrate spikes.

6. The alternate use of cosmogenic isotopes such as [10]Be to study past SPEs, advocated by many, is currently being accomplished by other researchers [*McCracken and Beer,* 2015], and they have identified the 4 of the 5 same events in the 1940-1956 time interval as the impulsive nitrate events identified in Figures 1 and 7 of *Smart et al*. [2014].  Only the GLE of 25 July 1946 was not captured in the [10]Be data. This is a strong vindication of the *Smart et al*. [2014] identification of these events in the ice core data.

7. High fluence, hard spectrum, winter events such as the 23 February 1956 SPE produce enough nitrate in mid-stratosphere down through the troposphere to account for ice core nitrate spikes and suggest that polar ice cores may be useful archives of such events covering the last several millennia. To our knowledge, this is the first study to obtain this result.  All previous simulations have concluded that nitrate deposition values in Greenland and Antarctica from SPEs were far too low to appear above the background.

As there is current [14]C data which may be produced by extremely high fluence past SPEs [*Miyake et al.,* 2013 and references therein; *Melott and Thomas*, 2012; *Usoskin et al.,* 2013; *Thomas et al*., 2013], combining cosmogenic isotope-based estimates [*Beer et al.,* 2011; *Kovaltsov et al*., 2014] with nitrate data may give some insight into the spectrum of hard events (which pose the greatest risk to aircraft and spacecraft) in the pre-technological past, and provide useful statistics on such events for the future. As an example, the non-detection of cosmogenic isotopes for the Carrington Event [*Usoskin and Kovaltsov* 2012; *McCracken and Beer* 2015] tells us that if nitrate spikes were to be found in future studies, the combination of this with the isotope data would constrain this event to be very high fluence, e.g. much greater than 1972 and sufficient to penetrate to the lower atmosphere, with a rather soft spectrum.

The probable medieval SPEs of 775 AD and 994 AD [*Miyake et al.,* 2013; *Melott and Thomas*, 2012; *Usoskin et al.*, 2013; *Ding et al.*, 2015; *Mekhaldi et al.,* 2015] are of concern. The 775 AD event may have been 25 - 50 times stronger than that of 23 February 1956 [*Usoskin et al.*, 2013] and could pose a threat to our technological civilization if it occurred today. There is considerable uncertainty in the spectrum and fluence of this event, but evidence to date suggests that it was likely very hard and high fluence [*Thomas et al.,* 2013; *Mekhaldi et al.,* 2015]. As an example, we consider that the 775 AD event might resemble that of 1956 scaled up conservatively by a factor of 10. As an exercise in what might be found in high-resolution data, we show in Figure 7 the GISP2H ice core data [*Dreschhoff and Zeller*, 1998] using these assumptions. If as strong as suggested, both 775 AD and 994 AD should be clearly detectable as large nitrate spikes above background in the ice cores. We note, however, a recent study [*Ding et al.*, 2015] indicates there may have been 3 smaller events that occurred in northern hemisphere summer 775-776 AD. If so, it suggests conditions would be more favorable for detecting this event in Antarctica than at Summit and that three smaller nitrate spikes would be less prominent.

*Smart et al.* [2014, 2016] showed that fine resolution is needed to see such events, yet nitrate analyses at sufficient resolution for these periods are currently lacking. Tree ring and coral $^{14}$C analyses have provided other stunning examples of the value in analysis at increased temporal resolution. We argue that additional fine-resolution ice core nitrate studies coupled with conductivity and multi species analysis for determining interferences from non SPE sources in both polar regions, covering the last two millennia, could shed important light on these events and should be undertaken. We urge consideration of such examination of ice cores from that period.

## 7. Acknowledgments

We are grateful for helpful comments from M. Shea, D. Smart. We thank the referees, whose comments improved the manuscript. I. Usoskin provided SPE-ionization data and A. Tylka provided spectral information on SPE events. ALM and BCT were supported by NASA grant NNX14AK22G. Initial creation of the tables of *Atri et al.* [2010] which were used was supported in part by the National Science Foundation through TeraGrid resources provided by the National Center for Supercomputing Applications. Additional computation time was provided by the High Performance Computing Environment (HiPACE) at Washburn University; thanks to Steve Black for assistance. We also acknowledge G.A.M. Dreschhoff for providing the detailed GISP2-H core data used in Figures 4 and 7. We thank Keri Nicoll for sharing the balloon measurement data.

## 9. Figures

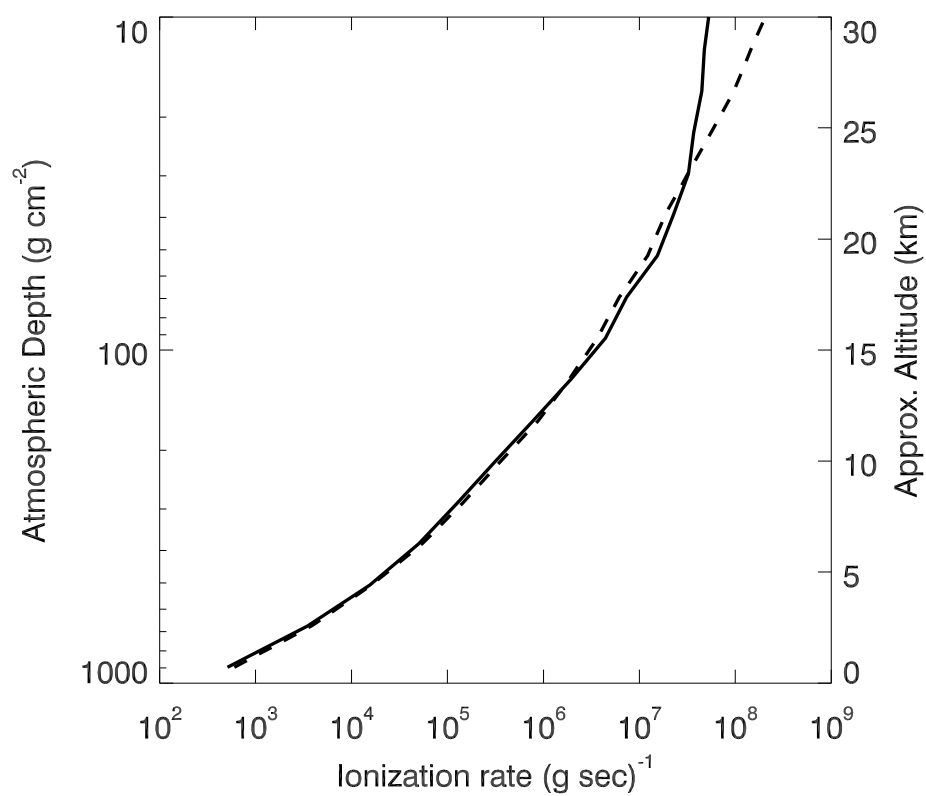

**Figure 1.**
Comparison of ionization rate profile generated using our model for primaries of energy greater than 300 MeV (solid line) and that presented in Figure 2B of *Usoskin et al.* [2011] (dashed line) for the 20 January 2005 SPE, assuming 1 day duration. The greatest difference is at the highest altitude (30 km), probably because *Usoskin et al.* [2011] included lower-energy primaries.

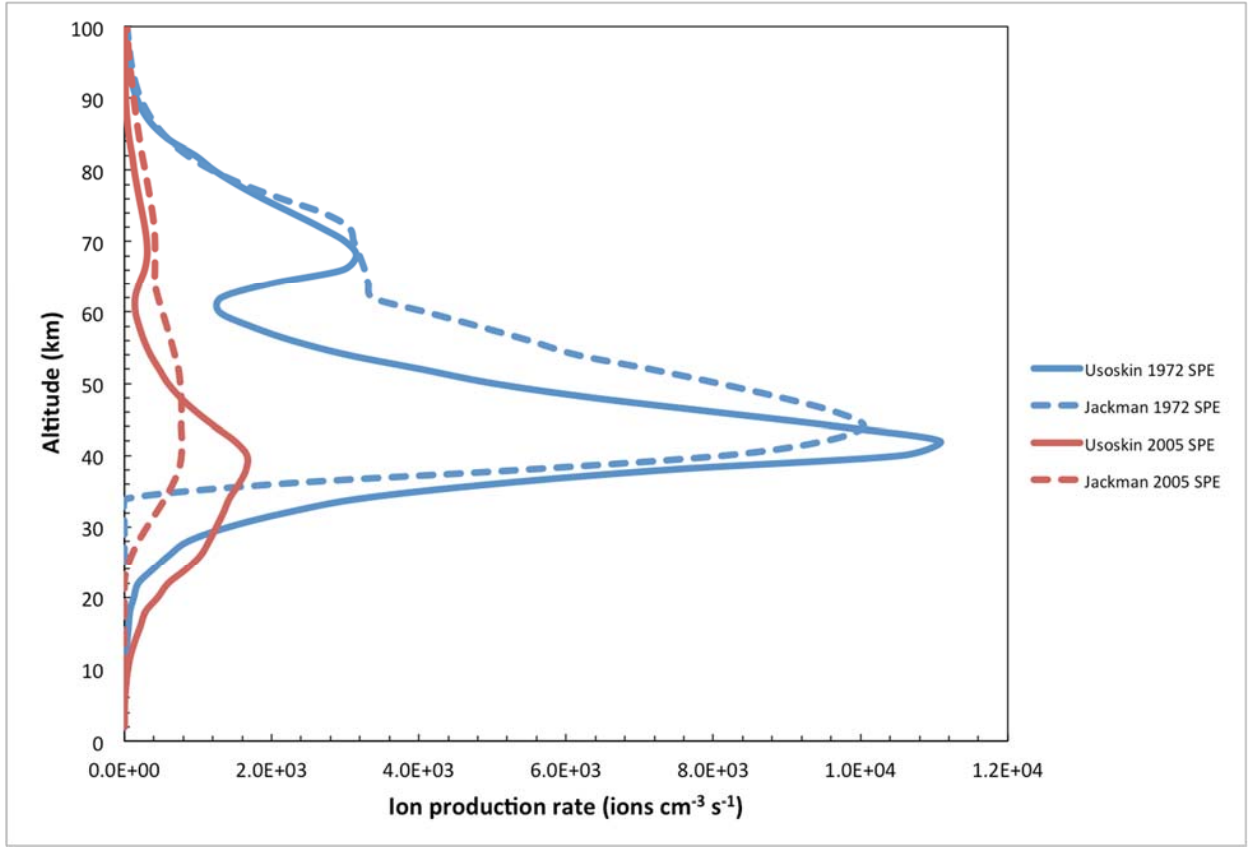

**Figure 2**

Comparison of ionization rate profiles for the 3-4 August 1972 SPE and the 20 January 2005 SPE, computed using two different methods: a low energy analytic-only method, data from C. Jackman (dashed lines); and a full air-shower treatment, data from I. Usoskin (solid lines). Note that the analytic-only SPE profiles significantly underestimate ionization rates in the mid to lower stratosphere. The disagreement is even more pronounced for the hard spectrum (but lower fluence) 2005 event.

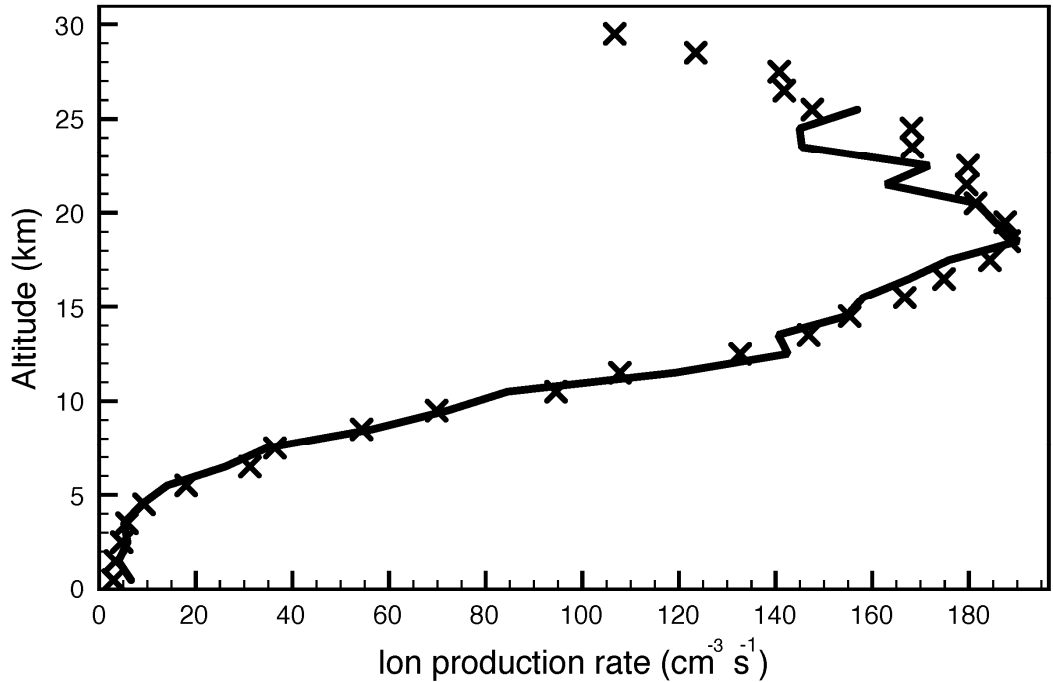

**Figure 3.**

Results of a balloon experiment described in *Nicoll and Harrison* [2014] which was launched on
April 11, 2013 from Reading, UK, compared with results of our air shower computations. A
proton spectrum with -2.4 spectral index was applied and added to the ambient ionization rate
(assuming the rigidity for 47.5 N at Reading, UK) measured by the balloon experiments. The
data was provided to us by Keri Nicoll. The measured ionization profile during the event is
shown as a solid line and compared against our calculated values (X marks). The maximum
percentage difference between the two is 16% at 24 km. We note that in the *Nicoll and Harrison*
[2014] Figure 2, right panel, the excess ionization due to the SPE peaks at 18 km and reaches
down to 5 km.

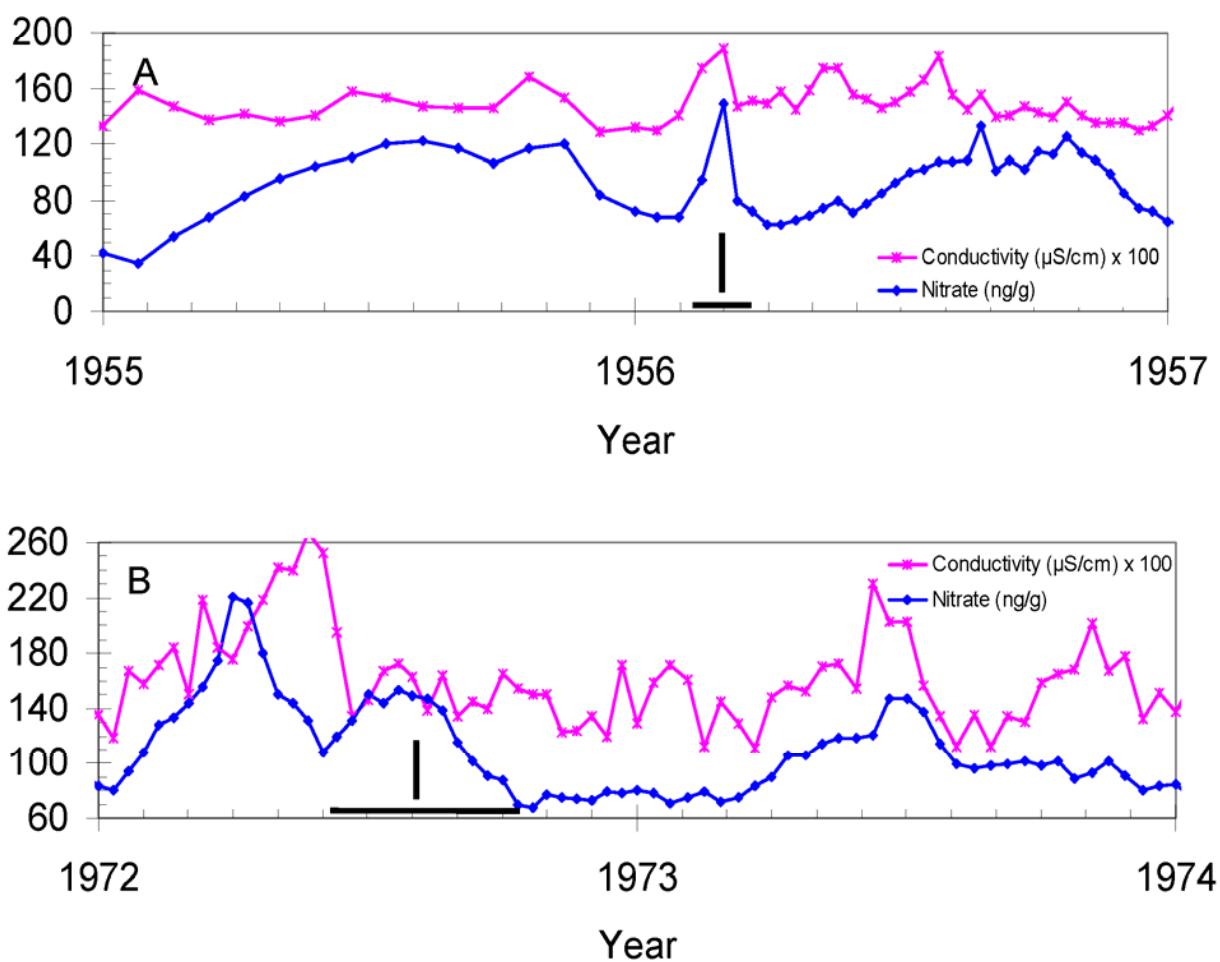

**Figure 4.**

The nitrate and conductivity records for two time periods in the GISP2H ice core, as discussed in the text. (A) The nitrate and conductivity peaks in early 1956 (indicated by horizontal bar, which runs between the adjacent nitrate minima) coincide with a hard-spectrum SPE and coincident GLE event on 23 February (time indicated approximately by the vertical bar). The area under the nitrate peak (comprised of 4 data points) and minus the local background (the average of the two adjacent end points and local minima) corresponds to a deposition of 116 ng $NO_3$ $cm^{-2}$ (see text). (B) The soft-spectrum SPE on 3-4 August 1972 (vertical bar) does not produce a nitrate peak clearly distinguishable above background here, though the secondary peak (indicated by horizontal bar) is a possible candidate.

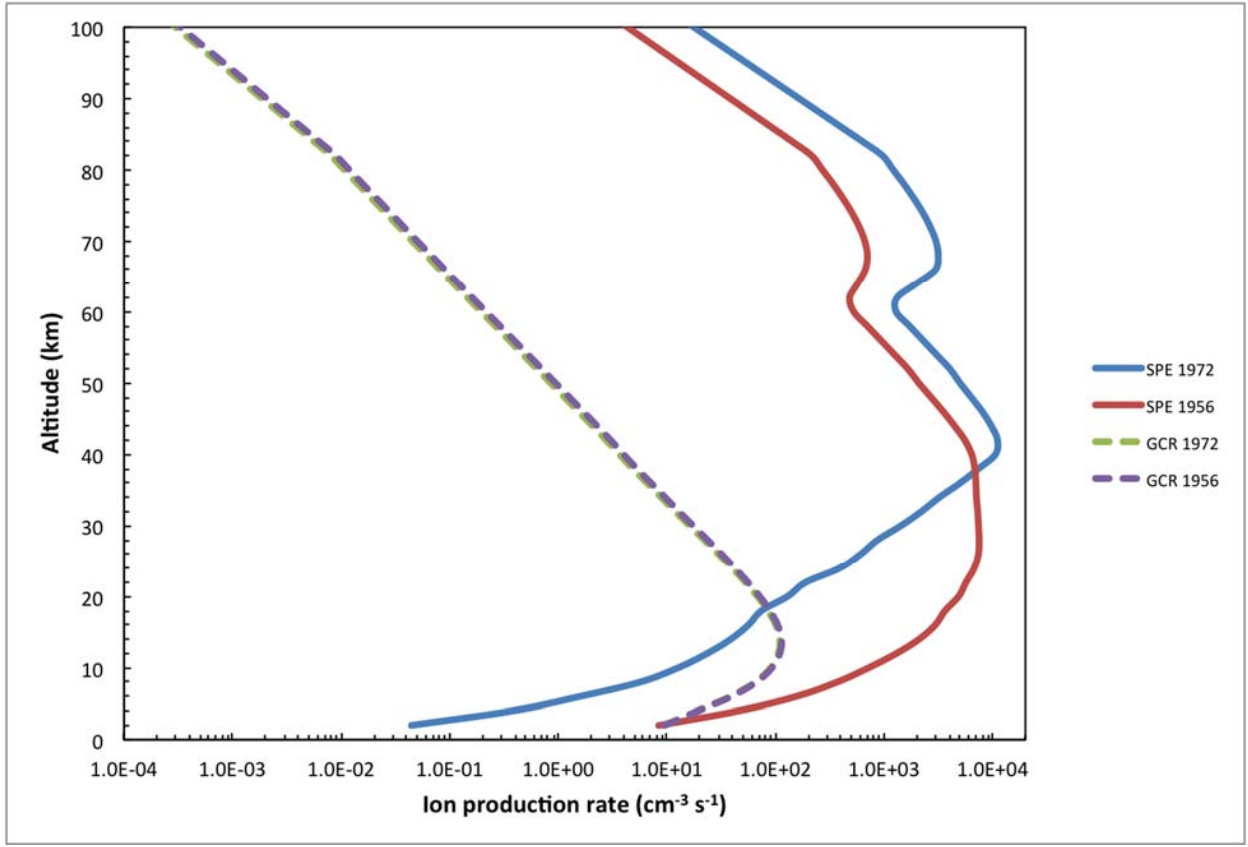

**Figure 5.**

Ionization rate profiles (provided by I. Uoskin). Note the logarithmic x-axis in contrast to the linear scale of Figure 2. The dashed lines are the background ionization rates (nearly identical) in the atmosphere due to Galactic Cosmic Rays in February 1956 and August 1972. The solid lines are ionization rates assuming a 1-day event duration. The red solid line shows ionization due to the 23 February 1956 SPE. Note that there is considerable ionization below 20 km, which is not shown or included in many computations. The blue solid line shows ionization due to the 3-4 August 1972 SPE, which does not produce a sharp peak clearly distinguishable above the background. This had a larger fluence of protons with energy above 30 MeV, but a softer spectrum than the 1956 event. Consequently, there is more ionization at high altitudes than in 1956, but much less at lower altitudes.

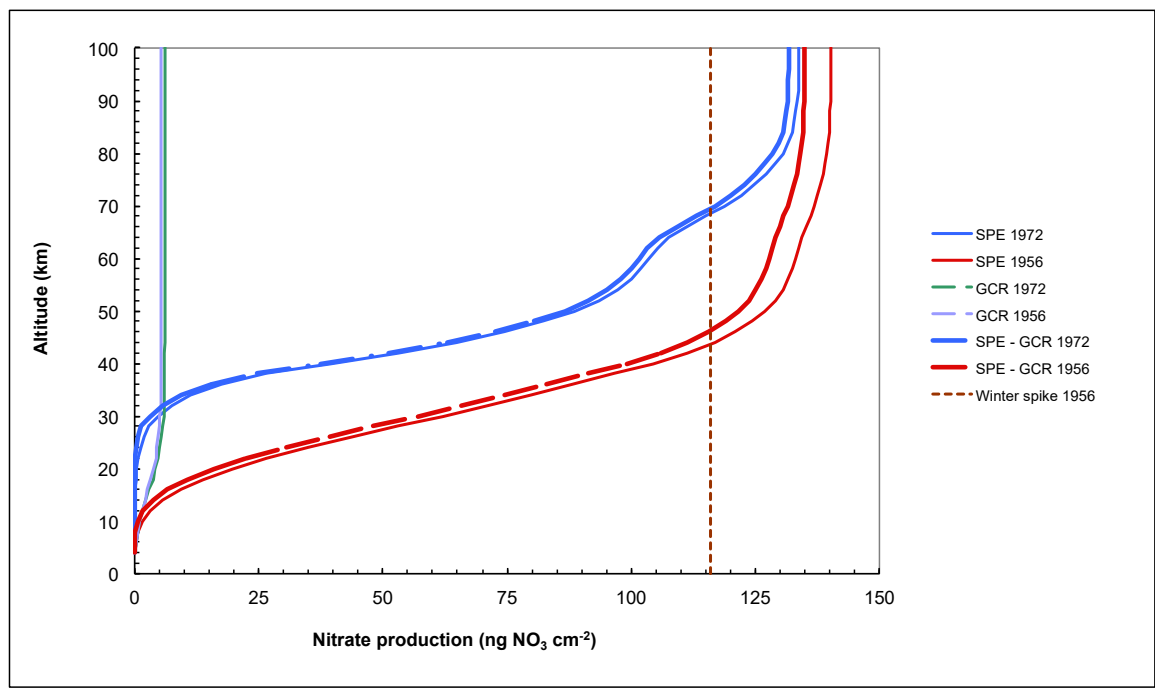

**Figure 6**

Integral nitrate production, i.e. summed from the surface (at 3.2 km elevation appropriate for Summit, Greenland) to the altitude indicated, computed from modeled ionization as described in Section 4. The profiles for 1972 and 1956 presented here are results of modeling using the 3-4 August 1972 and 23 February 1956 SPE spectra, the GCR ionization appropriate for those times (see Section 4), and the difference of the SPE and GCR contributions. The SPE-GCR values (dashed blue and red curves) give the enhancement over GCR background assuming a normal GCR flux (dashed green and purple curves). The SPE-only values (solid blue and red curves) give the enhancement in the case of no GCR background (ie. a complete Forbush decrease). The intersection of the Feb 1956 SPE and SPE - GCR nitrate production curves with the dashed vertical line (116 ng $NO_3$ cm$^{-2}$) at 44-46 km corresponds to the GISP2H ice core winter nitrate spike deposition above background and shows that nitrate production from 3.2 - 45 km by the 23 February 1956 SPE and subsequent deposition over 2 - 4 months is sufficient to explain the spike. Note that half of the nitrate is produced below 30 km, where deposition times are on the order of 1 month. As can be seen, the assumption of a Forbush decrease does not make a significant difference in sum values at the altitudes discussed in the text.

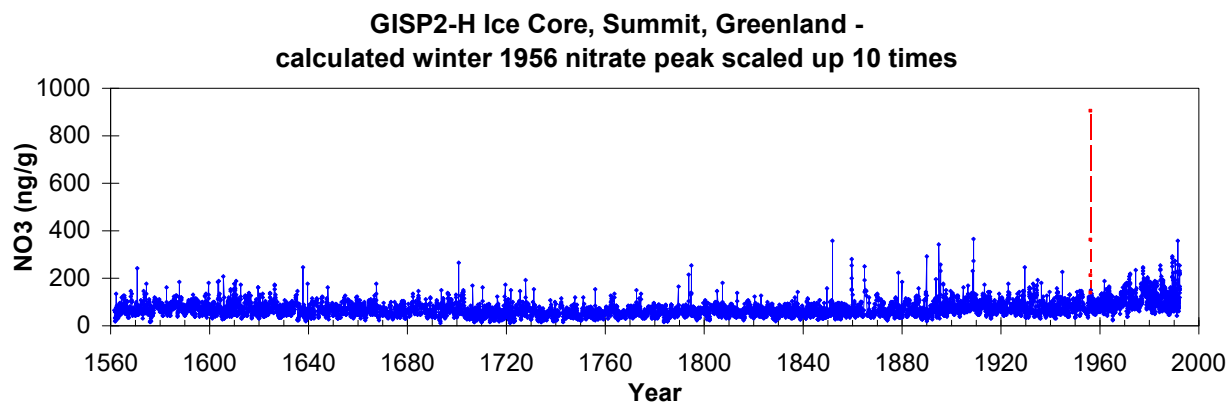

**Figure 7.**

The GISP2H ice core data [*Dreschhoff and Zeller*, 1998], but with the nitrate enhancement of the 1956 SPE (dashed portion of curve) scaled up by a factor of ten. It is intended as an example of the kind of information that may exist for events such as those in 775 and 993 AD, which were already found in $^{14}$C records from tree rings and corals but have not yet been sought systematically (i.e. at fine resolution) in the nitrate record in ice cores.  It should be noted that evidence suggesting several smaller events in 775 AD may have occurred (instead of one large one) and in northern hemisphere summer [*Ding et al.*, 2015] would greatly reduce the size of any resulting nitrate spikes at Summit accordingly.  It also suggests that Antarctica would be the place to look for nitrates associated with this event.  Comparing nitrate, $^{14}$C, and $^{10}$Be may enable estimates of the spectral shape of such events.